\documentstyle[12pt,graphicx]{article}
\title{Holographic hessence models }

\author{\small Wen Zhao\\
        \small Department of Applied Physics, Zhejiang University of Technology, \\
        \small Hangzhou, Zhejiang, People's Republic of China }
 \date{}

\begin{document}
\maketitle
\baselineskip=19truept
\def\vek{\vec{k}}

\newcommand{\be}{\begin{equation}}
\newcommand{\ee}{\end{equation}}
\newcommand{\ba}{\begin{eqnarray}}
\newcommand{\ea}{\end{eqnarray}}
\renewcommand{\H}{{\cal H}}
\renewcommand{\L}{{\cal L}}

\sf\small

\begin{center}
\Large  Abstract
\end{center}
\begin{quote}
 { \small

We discuss the evolution of holographic hessence model, which
satisfies the holographic principle and can naturally realizes the
equation of state crossing $-1$. By discussing the evolution of
the models in the $w-w'$ plane, we find that, if $c\geq1$,
$w_{he}\geq-1$ and $\dot{V}<0$ keep for all time, which are
quintessence-like. However, if $c<-1$, which mildly favors the
current observations, $w_{he}$ evolves from $w_{he}>-1$ to
$w_{he}<-1$, and the potential is a nonmonotonic function. In the
earlier time, the potential must be rolled down, and then be
climbed up. Considered the current constraint on the parameter
$c$, we reconstruct the potential of the holographic hessence
model.

 }
\end{quote}




e-mail: wzhao7@mail.ustc.edu.cn

\baselineskip=17truept

\small

\section{Introduction}

Numerous and complementary cosmological observations indicate that
the expansion of the universe is undergoing cosmic acceleration at
the present time\cite{observation}. This cosmic acceleration is
viewed as due to a mysterious dominant component, dark energy,
with negative pressure. The combined analysis of cosmological
observations suggests that the universe is spatially flat, and
consists of about $70\%$ dark energy, $30\%$ dust matter (cold
dark matter plus baryons), and negligible radiation. Although we
can affirm that the ultimate fate of the universe is determined by
the feature of dark energy, the nature of dark energy as well as
its cosmological origin remain enigmatic at present. Explanations
have been sought within a wide range of physical phenomena,
including a cosmological constant, exotic
fields\cite{quint,phantom,k,quintom,vector}, a new form of the
gravitational equation\cite{gt}, etc. Recently, a new model
stimulated by the holographic principle has been put forward to
explain the dark energy\cite{Hsu:2004ri,Li:2004rb}. According to
the holographic principle, the number of degrees of freedom of a
physical system scales with the area of its boundary. In the
context, Cohen et al\cite{cohen} suggested that in quantum field
theory a short distant cutoff is related to a long distant cufoff
due to the limit set by formation of a black hole, which results
in an upper bound on zero-point energy density. In line with this
suggest, Hsu and Li\cite{Hsu:2004ri,Li:2004rb} argued that this
energy density could be views as the holographic dark energy
satisfying

\be \rho_{de}=3c^2M_P^2L^{-2}~,\label{de} \ee where $c$ is a
numerical constant, and $M_P\equiv 1/\sqrt{8\pi G}$ is the reduced
Planck mass. If we take $L$ as the size of the current universe,
for instance the Hubble scale $H^{-1}$, then the dark energy
density will be close to the observed data. However,
Hsu\cite{Hsu:2004ri} pointed out that this yields a wrong equation
of state for dark energy. Li\cite{Li:2004rb} subsequently proposed
that the IR cut-off $L$ should be taken as the size of the future
event horizon

\be L=R_{eh}(a)=a\int_t^\infty{d\tilde{t}\over
a(\tilde{t})}=a\int_a^\infty{d\tilde{a}\over
H\tilde{a}^2}~.\label{eh} \ee Then the problem can be solved
nicely and the holographic dark energy model can thus be
constructed successfully. The holographic dark energy scenario may
provide simultaneously natural solutions to both dark energy
problems as demonstrated in Ref.\cite{Li:2004rb}. The only
undetermined parameter $c$ should be fixed by the observations. If
$c\leq1$, which satisfies the original bound $L^3\rho_{de}\leq
LM_p^2$, the equation of state (EOS) of dark energy evolves from
the state of $w>-1$ to $w<-1$, and the critical state of $w=-1$
must be crossed. If $c>1$, the EOS of dark energy keeps
$w>-1$\cite{Li:2004rb}, which naturally avoided the cosmic big
rip. However, the original bound $L^3\rho_{de}\leq LM_p^2$ will be
violated. Since the model we discuss here is only a
phenomenological framework and it is unclear whether it is
appropriate to tightly constrain the value of $c$ by means of the
analogue to the black hole. As a matter of fact, the possibility
of $c>1$ has been seriously dealt with and a modest value of $c$
larger than one could be favored in the literature\cite{c>1}. In
this paper, we consider the general case with $c$ as a free
parameter.

For a kind of realized dark energy model, the feature of EOS
crossing $-1$ can not be realized by the simple quintessence,
phantom, or k-essence\cite{-1}. The quintom is one of the simplest
models with EOS crossing $-1$, which is the combination of a
quintessence $\phi_1$ and a phantom $\phi_2$. The hessence is a
kind of simple quintom\cite{hessence,zhao}, which has the
lagrangian density
 \be\label{8}
 \L_{he}=\frac{1}{2}(\partial_{\mu}\phi_1)^2-\frac{1}{2}
 (\partial_{\mu}\phi_2)^2-V(\phi_1^2-\phi_2^2)~,
 \ee
where the potential function $V(\phi_1^2-\phi_2^2)$ is free for
the models. Different choice of $V$ follows a different evolution
of the universe. In Ref.\cite{brane}, the authors found that this
kind of models may be the local effective approximation of the
D3-brane Universe. In Ref.\cite{zhao}, we have proved that the
evolution of potential function can be exactly determined by the
EOS of hessence $w_{he}(z)$ and its evolution $w_{he} '(z)$. If
considered the holographic constraint in Eq.(\ref{de}), the EOS of
the hessence can be exactly determined for a fixed $c$. So the
potential function for the holographic hessence only depends on
the parameter $c$. In this letter, we first discuss the evolution
of the EOS and potential of the holographic hessence models for
the different $c$. Then considered the constraint on $c$ from the
current observations, we reconstruct the potential function of
holographic hessence models.

\section{Holographic hessence models}

We consider the action
 \be\label{s}
 S=\int d^4x\sqrt{-g}\left(-\frac{\cal R}{16\pi G}+\L_{he}+\L_m\right),
 \ee
where $g$ is the determinant of the metric $g_{\mu\nu}$, $\cal R$
is the Ricci scalar, $\L_{he}$ and $\L_m$ are the lagrangian
densities of the hessence dark energy and matter, respectively.
The lagrangian density of hessence is in Eq.(\ref{8}). One can
easily find that this lagrangian is invariant under the
transformation
 \be
 \phi_1\rightarrow\phi_1\cosh(i\alpha)-\phi_2\sinh(i\alpha)~,
 \ee
 \be
 \phi_2\rightarrow-\phi_1\sinh(i\alpha)+\phi_2\cosh(i\alpha)~,
 \ee
where $\alpha$ is constant. This property makes one can rewrite
the lagrangian density (\ref{8}) in another form
 \be\label{he}
 \L_{he}=\frac{1}{2}
 \left[(\partial_{\mu}\phi)^2-\phi^2(\partial_{\mu}\theta)^2\right]-V(\phi)~,
 \ee
where we have introduced two new variables $(\phi,~\theta)$, i.e.
 \be
 \phi_1=\phi\cosh\theta~,~~~~~~\phi_2=\phi\sinh\theta~.
 \ee
Consider a spatially flat FRW (Friedmann-Robertson-Walker)
universe with metric
 \be
 ds^2=dt^2-a^2 (t) \gamma_{ij}dx^idx^j~,
 \ee
where $a(t)$ is the scale factor, and $\gamma_{ij}=\delta^i_j$
denotes the flat background space. Assuming $\phi$ and $\theta$
are homogeneous, from the action in (\ref{s}), we obtain the
equations of motion for $\phi$ and $\theta$
 \be\label{1}
 \ddot{\phi}+3H\dot{\phi}+\phi\dot{\theta}^2+dV/d\phi=0~,
 \ee
 \be\label{2}
 \phi^2\ddot{\theta}+(2\phi\dot{\phi}+3H\phi^2)\dot{\theta}=0~,
 \ee
where $H\equiv\dot{a}/a$ is the Hubble parameter, an overdot
denotes the derivatives with respect to cosmic time. Eq.(\ref{2})
implies
 \be
 Q=a^3\phi^2\dot{\theta}={\rm const}~,
 \ee
which is associated with the total conserved charge within the
physical volume due to the internal symmetry\cite{hessence}. This
relation turns out
 \be
 \dot{\theta}=\frac{Q}{a^3\phi^2}~.
 \ee
Substituting this into Eq.(\ref{1}), we can rewrite the kinetic
equation as
 \be\label{kinetic}
 \ddot{\phi}+3H\dot{\phi}+\frac{Q^2}{a^6\phi^3}+\frac{dV}{d\phi}=0~,
 \ee
which is equivalent to the energy conservation equation of the
hessence $\dot{\rho}_{he}+3H(\rho_{he}+p_{he})=0$. The pressure,
energy density and the EOS of the hessence are
 \be\label{24}
 p_{he}=\frac{1}{2}\dot{\phi}^2-\frac{Q^2}{2a^6
 \phi^2}-V(\phi)~,~~~~~~~
 \rho_{he}=\frac{1}{2}\dot{\phi}^2-\frac{Q^2}{2a^6 \phi^2}+V(\phi)~,
 \ee
 \be\label{25}
 w_{he}=\left.\left[\frac{1}{2}\dot{\phi}^2-\frac{Q^2}{2a^6
 \phi^2}-V(\phi)\right]\right/
 \left[\frac{1}{2}\dot{\phi}^2-\frac{Q^2}{2a^6
 \phi^2}+V(\phi)\right]~,
 \ee
respectively. It is easily seen that $w_{he}\geq-1$ when
$\dot{\phi^2}\geq Q^2/(a^6\phi^2)$, while $\omega_{he}\leq-1$ when
$\dot{\phi^2}\leq Q^2/(a^6\phi^2)$. The transition occurs when
$\dot{\phi^2}=Q^2/(a^6\phi^2)$. In the case of $Q\equiv0$, the
hessence becomes the quintessence model. From the expression of
EOS of hessence, we can find it is only dependant of the potential
function $V(\phi)$. If $V(\phi)$ is determined, $w$ is also
determined. On the contrary, if $w(z)$ is fixed, the potential
function $V(\phi)$ also can be solved. Here we consider the
holographic hessence models, which satisfies the holographic
constraint in Eq.(\ref{de}). Consider now a spatially flat FRW
universe with matter component $\rho_{m}$ (including both baryon
matter and cold dark matter) and holographic hessence component
$\rho_{he}$. The Friedmann equation reads
\begin{equation}
3H^2M_p^2=\rho_{m}+\rho_{he}~,
\end{equation} or equivalently,
\begin{equation}
{H^2\over H_0^2}=\Omega_{m0}a^{-3}+\Omega_{he}{H^2\over
H_0^2}~.\label{Feq}
\end{equation}
Combining the definition of the holographic dark energy (\ref{de})
and the definition of the future event horizon (\ref{eh}), we
derive
\begin{equation}
\int_a^\infty{d\ln \tilde{a}\over H\tilde{a}}={c\over
Ha\sqrt{\Omega_{he}}}~.\label{rh}
\end{equation} We notice that the Friedmann
equation (\ref{Feq}) implies
\begin{equation}
{1\over Ha}=\sqrt{a(1-\Omega_{he})}{1\over
H_0\sqrt{\Omega_m^0}}~.\label{fri}
\end{equation} Substituting (\ref{fri}) into (\ref{rh}), we get easily the dynamics satisfied
by the dark energy, i.e. the differential equation about the
fractional density of dark energy,
\begin{equation}
\Omega'_{he}=\Omega_{he}(1-\Omega_{he})\left(1+{2\over
c}\sqrt{\Omega_{he}}\right),\label{deq}
\end{equation}
where the prime denotes the derivative with respect to $\ln a$.
This equation describes behavior of the holographic dark energy
completely, and it can be solved exactly. It is easy to prove
that, this equation has only one steady attractor solution
 \be
 \Omega_{he}=1.
 \ee
In the solution, the hessence is dominant in the universe, and the
component of matter is negligible.

Important observables to reveal the nature of dark energy are the
EoS $w$ and its time derivative in units of Hubble time $w'$. The
SNAP mission is expected to observe about $2000$ SNIa each year,
over a period of three years. Most of these SNIa are at the
redshift $z\in[0.2, ~1.2]$. The SNIa plus weak lensing methods
conjoined can determine the present equation of state ratio,
$\omega_0$, to $5\%$, and its time variation, $\omega'$, to $0.11$
\cite{snap}. It has a powerful ability to differentiate the
various dark energy models. From the energy conservation equation
of the holographic hessence, the EOS of the dark energy can be
given \cite{Li:2004rb}
\begin{equation}\label{ww}
w_{he}=-1-{1\over 3}{d\ln\rho_{he}\over d\ln a}=-{1\over
3}\left(1+{2\over c}\sqrt{\Omega_{he}}\right)~,
\end{equation}
and its evolution is
 \be\label{wwp}
 w_{he}'=-\frac{\sqrt{\Omega_{he}}}{3}(1-\Omega_{he})\left(1+\frac{2}{c}\sqrt{\Omega_{he}}\right).
 \ee
It can be seen clearly that the equation of state of the
holographic dark energy evolves dynamically and satisfies
$-(1+2/c)/3\leq w_{he}\leq -1/3$ due to $0\leq\Omega_{he}\leq 1$.
The parameter $c$ plays a significant role in this model. If one
takes $c=1$, the behavior of the holographic dark energy will be
more and more like a cosmological constant with the expansion of
the universe, such that ultimately the universe will enter the de
Sitter phase in the far future. As is shown in\cite{Li:2004rb}, if
one puts the parameter $\Omega_{he0}=0.73$ into (\ref{ww}), then a
definite prediction of this model, $w_{he0}=-0.903$, will be
given. On the other hand, if $c<1$, the holographic dark energy
will exhibit appealing behavior that the equation of state crosses
the ``cosmological-constant boundary'' (or ``phantom divide'')
$w=-1$ during the evolution. This kind of dark energy is referred
to as ``quintom'' \cite{quintom} which is slightly favored by
current observations \cite{obs1}. If $c>1$, the equation of state
of dark energy will be always larger than $-1$ such that the
universe avoids entering the de Sitter phase and the Big Rip
phase. Hence, we see explicitly, the value of $c$ is very
important for the holographic dark energy model, which determines
the feature of the holographic hessence as well as the ultimate
fate of the universe.

Now, we discuss the dark energy models in the $w-w'$ plane, which
clearly shows the evolution character of the dark energy. The
simplest model, cosmological constant, has the effective state of
$w=-1$ and $w'=0$, which corresponds to a fixed point in the
$w-w'$ plane. Generally, the dynamics model of dark energy shows a
line in this plane, which describes the evolution of its
EOS\cite{wwp}. The simple quintessence has the state of $w\geq-1$,
which only occupies the region of $w'>-3(1+w)(1-w)$. The phantom
field ($w\leq-1$) occupies the region of $w'<-3(1+w)(1-w)$. The
evolution of hessence in the $w-w'$ plane is discussed in
Ref.\cite{zhao}. Here we brief it as below. From the kinetic
equation (\ref{kinetic}), one can get
 \be\label{32}
 1+\frac{1}{6}\frac{d\ln x}{d\ln
 a}=-\frac{1}{3HV}\frac{\dot{V}}{1+w_{he}}~,
 \ee
where $a$ is the scale factor, and we have set the present scalar
factor $a_0=1$. The function $x$ is defined by
 \be
 x\equiv\left|\frac{1+w_{he}}{1-w_{he}}\right|=\left|\frac{\frac{1}{2}
 \dot{\phi}^2-\frac{Q^2}{2a^6\phi^2}}{V}\right|~,
 \ee
and
 \be
 \frac{d\ln x}{d\ln a}=\frac{2w_{he}'}{(1+w_{he})(1-w_{he})}~.
 \ee
This equation can be rewritten as
 \be\label{37}\left.
 \left(1+\frac{w_{he}'}{3(1+w_{he})(1-w_{he})}\right)\right/\left(\frac{2\dot{V}}{3H(1+w_{he})\rho_{he}}\right)=-\frac{\rho_{he}}{2V}<0~.
 \ee
which follows that
 \be\label{FV}
 F\dot{V}<0,
 \ee
where we have defined $F\equiv w_{he}'+3(1+w_{he})(1-w_{he})$. So
the $w-w'$ plane is divided into four parts

~~~~~~~~~~~~~~{\bf I}:~~~~~~ $F>0$ ~~\&~~
 $w>-1$;~~~~~~~{\bf II}:~~~~~~~$F>0$ ~~\&~~
 $w<-1$;

~~~~~~~~~~~{\bf III}:~~~~~~ $F<0$ ~~\&~~
 $w<-1$;~~~~~~{\bf IV}:~~~~~~ $F<0$ ~~\&~~
 $w>-1$.\\
This can be seen clearly in Figure 1. From Eq.~(\ref{FV}), one can
easily find that $\dot{V}<0$ is satisfied in Region \emph{I} and
\emph{II} (rolling-down region), the field rolling down the
potential, and $\dot{V}>0$ is satisfied in Region \emph{III} and
\emph{IV} (climbing-up region), the field climbing up the
potential. So from the value of the function $w'+3(1-w)(1+w)$
being positive or negative, one can immediately judge how the
field evolves at that time. Toward the holographic hessence
models, from the equations (\ref{ww}) and (\ref{wwp}), we have
 \be
 F=\frac{1}{3}\left(\frac{2}{c}\Omega_{he}^2+\Omega_{he}^{\frac{3}{2}}
 -\frac{4+2c}{c^2}\Omega_{he}-\frac{c+4}{c}\sqrt{\Omega_{he}}+8\right).
 \ee
For a fixed $c$, $F$ only depends on the value of $\Omega_{he}$.
So the evolution of potential function of hessence is also exactly
determined by the evolution of $\Omega_{he}$. It is easily to find
that this is a monotonic increasing function with the increasing
$\Omega_{he}$. In order to exactly determine this function, one
must numerically solve the Eq.(\ref{deq}) and get function
$\Omega_{he}(z)$. Here we only concern on that whether the
potential function of holographic hessence is monotonic. If $F<0$
is holden for all time, the potential is a monotonic increasing
function. But it is not same with the phantom models, since the
cosmological constant can be crossed in the this hessence. If
$F>0$ is holden for all time, the potential is a monotonic
decreasing function. Since $w=-1$ also can be crossed in this
case, the models are different from the simple quintessence. If
the state of $F=0$ is crossed in the evolution of hessence, the
potential function of hessence can not be a monotonic function.

Here we focus on the initial and finial states of the holographic
hessence, and investigate them in the $w-w'$ plane. In the initial
condition with $\Omega_{he}\rightarrow0$, one has
 \be
 (w_{he},w_{he}')\rightarrow\left(-\frac{1}{3},~0\right),
 \ee
 \be
 F=w_{he}'+3(1+w_{he})(1-w_{he})\rightarrow\frac{8}{3}>0,
 \ee
which is in the rolling-down region (Region \emph{I}), and
independent of value of $c$. So in any case of models, they evolve
from the region \emph{I}, which is quintessence-like. However, in
the finial stage with $\Omega_{he}\rightarrow1$, one has
 \be
 w_{he}\rightarrow-{1\over
3}\left(1+{2\over c}\right),~~~~w_{he}'\rightarrow0,
 \ee
and
 \be
 F=\frac{4}{3}\left(2-\frac{1}{c}-\frac{1}{c^2}\right).
 \ee
For the different choice of $c$, the finial value of $F$ is
different.
 \ba
 c&>&1,~~F>0~~~(~\rm in~~the~~rolling-down~~region~),\\
 c&=&1,~~F=0~~~(~\rm at~~the~~critical~~point~),\\
 c&<&1,~~F<0~~~(~\rm in~~the~~climbing-up~~region~).
 \ea
In Figure 1, we have plotted the evolution of three different
holographic hessence models with $c=1.5,1.0,0.5$, respectively,
where the arrows indicate the evolution direction of hessence with
the increasing value of $\Omega_{he}$ from $\Omega_{he}=0$ to
$\Omega_{he}=1$. From this figure, we can find that, if $c>1$ is
chosen, which violates the holographic constraint, the hessence is
in the Region \emph{I} (rolling-down region) for all time, so the
potential of hessence is a monotonic damping function, which is
similar the quintessence models. This is consistent to the
previous conclusion that holographic dark energy with $c>1$ can be
described by the quintessence fields. However, if the fixed $c$ is
smaller than $1$, which satisfies the holographic constraint, the
hessence must evolve from Region \emph{I} (rolling-down region) to
Region \emph{IV} (climbing-up region), and finally to Region
\emph{III} (climbing-up region). So the potential of hessence is
not a monotonic function. The field $\phi$ rolls down the
potential at earlier stage, and later it turns to climb up. With
the expansion of the universe, the state of $F=0$ must be crossed.
The EOS of hessence also turns from the region of $w>-1$ to that
of $w<-1$, and the state of $w=-1$ must be crossed. So dark energy
is quintom-like. We note that the time of $F=0$ is a little
earlier than that of $w=-1$. If $c=1$ is chosen, the holographic
hessence is also in Region \emph{I} for all time, and finally it
turns to the critical state of $(w,w')=(-1,0)$ with the expansion
of the universe, and the universe is an exact de Sitter expansion.

 \begin{figure}
 \centerline{\includegraphics[width=10cm]{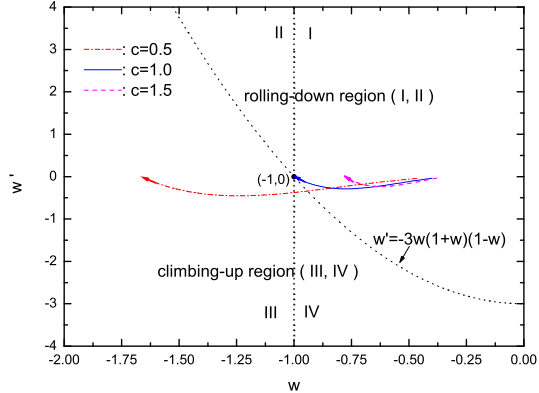}}
 \caption{\small The holographic hessence models evolve in the $w-w'$ plane, where we have considered
 three models with $c=0.5,1.0,1.5$, respectively. The arrows denote the
 evolution direction of the models with the expansion of the universe.  }
 \end{figure}

\section{Reconstruct the holographic hessence models}

From the previous discussion, we find the value of the parameter
$c$ should be fixed by the cosmological observations. This has
been discussed by a number of authors\cite{obs1}. In the recent
work\cite{zx}, the authors have constrained the holographic dark
energy by the current observations of SNIa (Type Ia Supernova),
CMB (cosmic microwave background radiation), and BAO (baryon
acoustic oscillation). If setting $c$, $\Omega_{m0}$ and $H_0$ as
the free parameters, and only using the up-to-date gold sample of
SNIa consisted of $182$ data\cite{snia}, the author found that the
best-fit for the analysis of gold sample of SNIa happens at
 \be\label{fit1}
 c=0.37,~~\Omega_{m0}=0.43,~~h=0.64.
 \ee
By choosing $h=0.64$, the $1\sigma$ fit values for the parameters
are:
 \be
 c=0.37^{+0.56}_{-0.21},~~ \Omega_{m0}=0.43^{+0.08}_{-0.14}.
 \ee
It is obvious that the SNIa data alone seem not sufficient to
constrain the holographic dark energy models strictly. The
confidence region of $c$ is very large, and the best fit of
$\Omega_{m0}$ is evidently different from other
constraint\cite{seljak}. In the previous work\cite{0506310}, the
authors found that the holographic dark energy model is very
sensitive to the value of the present Hubble parameter $h$. So it
is very important to use other results of CMB and LSS (large-scale
structure), which are observational quantities irrelevant to $h$
as a complement to SNIa data. The authors considered the recent
observations on the CMB shift parameter $R=1.70\pm0.03$\cite{cmb}
and the measurement of BAO peak in the distribution of SDSS
luminous red galaxies\cite{bao}. From the constraints of the
combination of SNIa, CMB and BAO, and considered the prior
$h=0.72\pm0.08$, which is got from the \emph{Hubble Space
Telescope Key Project} (\emph{HST})\cite{h}, the fit values for
model parameters with $1-\sigma$ errors are
 \be\label{fits}
 c=0.91^{+0.23}_{-0.19},~~\Omega_{m0}=0.29\pm0.03.
 \ee
It is clear that in the joint analysis the derive value for matter
density $\Omega_{m0}$ is very reasonable, which is important for
this model is the determination of the value of $c$.

As shown in \cite{0506310} and the discussion above, the
constraint of $h$ can evidently change the constraint result. In
order to show how strongly biased constraints can be derived from
a factitious prior on $h$, the author also considered a strong
$HST$ prior, fixing $h=0.72$. The constraint in equation
(\ref{fits}) becomes
 \be\label{fit3}
 c=0.42\pm0.05,~~\Omega_{m0}=0.24^{+0.02}_{-0.03}.
 \ee
We find that the confidence level contours get very evident
shrinkage and left-shift in the $c-\Omega_{m0}$ parameter-plane,
which also changes the evolution of the EOS parameter of the dark
energy and deceleration parameter of the universe\cite{zx}. These
all exactly consist with the previous results\cite{0506310}. We
also find that the constraint of $c$ in (\ref{fits}) and
(\ref{fit3}) are not overlapped, which is because that the
confidence level in these results are two low. If considered the
fit values for model parameters with $3-\sigma$ errors, the
conclusion will be much improved\cite{zx,0506310}.

However, if setting the Hubble constant as a free parameter in the
range of $(0.64,0.80)$, the constraint becomes
 \be\label{fit4}
 c=0.82^{+0.11}_{-0.13},~~\Omega_{m0}=0.28^{+0.03}_{-0.02}.
 \ee
which also have shrinkage and left-shift in the $c-\Omega_{m0}$
plane, comparing with the results in (\ref{fits}).

From these joint analysis, we can find that, though the
possibility of $c>1$ can not be excluded in one-sigma error range,
the possibility of $c<1$ is much more favored, which determined
that the dark energy is quintom-like, and the EOS crosses $-1$ at
some time.

 \begin{figure}
 \centerline{\includegraphics[width=10cm]{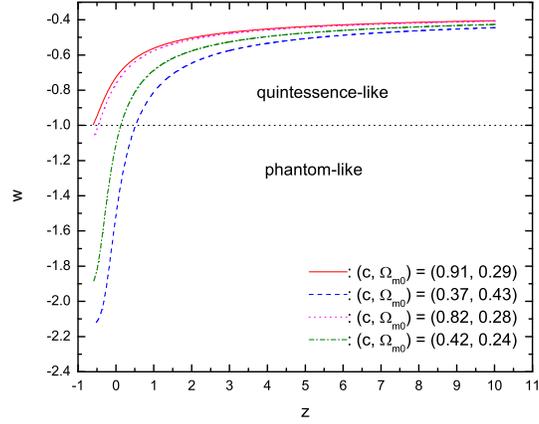}}
 \caption{\small From the observations, we solve the EOSs of the holographic hessence models.}
 \end{figure}

 \begin{figure}
 \centerline{\includegraphics[width=10cm]{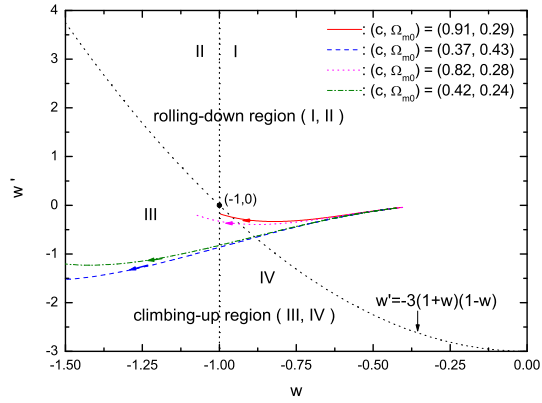}}
 \caption{\small The reconstructed holographic hessence models evolve in the $w-w'$ plane, where the arrows denote the
 evolution direction of the models with the expansion of the universe. }
 \end{figure}

From the differential equation (\ref{deq}), we can get the
evolution equation of $\Omega_{he}$ with the redshift $z$
 \be
 \frac{d\Omega_{he}}{dz}=-(1+z)^{-1}\Omega_{he}(1-\Omega_{he})\left(1+{2\over
c}\sqrt{\Omega_{he}}\right).
 \ee
For the determined parameters $c$ and
$\Omega_{he0}=1-\Omega_{m0}$, one can numerically solve this
equation and get $\Omega_{he}=\Omega_{he}(z)$. Inserting this into
equations (\ref{ww}) and (\ref{wwp}), we can get the EOS of the
holographic hessence $w=w(z)$ and its evolution $w'=w'(z)$. In
Figures 2 and 3  , we have plotted the EOS and its evolution of
the holographic hessence with best-fit parameters in the $w-z$ and
$w-w'$ planes. From Figure 2, we find that, in the earlier stage
of universe, the $w>-1$ holds for all cases, and the hessence are
quintessence-like. But the values of EOS decrease with time, and
they become phantom-like at present time for the case of $c=0.37$
and $c=0.42$. So the cosmological constant has been crossed. In
the cases of $c=0.91$ and $0.82$, though $w>-1$ is holden until
now, $w<-1$ will occur in the near future, and crossing the
cosmological constant is unavoidable. This feature determines that
this holographic dark energy can not be described by the
quintessence, phantom, k-essence, or Yang-Mills field
models\cite{-1,vector}. But in the hessence models, it can be
naturally and simply realized. From Figure 3, we find that, in the
earlier stage, the hessence models are all in Region \emph{I}
(rolling-down region). With the expansion of the universe, these
all models will cross the line with $F=0$ and enter into the
Region \emph{IV}, where although $w>-1$ is kept, the hessence
fields begin to clime up the potentials. At last, the hessence
models all cross the cosmological constant bound and stay in the
Region \emph{III}, where the hessences are phantom-like, and the
potentials are climbed up. So the potentials of these holographic
hessence are not a monotonic function, which is consistent to the
previous discussion.

With these solved EOSs, we can reconstruction the potential of the
holographic hessence models. Consider the FRW Universe, which is
dominated by the non-relativistic matter and a spatially
homogeneous hessence field $\phi$. The energy conservation
equation of the hessence field is
 \be
 \dot{\rho}_{he}+3H(\rho_{he}+p_{he})=0~,
 \ee
which yields
 \be\label{58}
 \rho_{he}(z)=\rho_{he0} \exp\left[3\int_0^z(1+w_{he})d\ln(1+\tilde{z})\right]\equiv
 \rho_{he0}E(z)~,
 \ee
where the subscript $0$ denotes the value of a quantity at the
redshift $z=0$ (present). From the expresses of the pressure and
energy density of the hessence, we get
 \be\label{55}
 V(\phi)=\frac{1}{2}\left(1-w_{he}\right)\rho_{he}~,
 \ee
 \be\label{56}
 \dot{\phi}^2=\frac{Q^2}{a^6\phi^2}+(1+w_{he})\rho_{he}~.
 \ee
Inserting the formula in these two equations, and after some
normal calculation, we get\cite{zhao}
 \be\label{63}
 \frac{d\tilde{\phi}}{dz}=\frac{\sqrt{3}}{(1+z)}
 \left[\frac{C(1+z)^6\tilde{\phi}^{-2}+(1+w_{he})E(z)}{r_0(1+z)^3+E(z)}\right]^{1/2}~,
 \ee
 \be\label{64}
 \tilde{V}[\phi]=\frac{1}{2}(1-w_{he})E(z)~,
 \ee
where $r_0\equiv\Omega_{m0}/\Omega_{he0}$ is the energy density
ratio of matter to hessence at present time, and the dimensionless
quantities are defined by
 \be\label{62}
 \tilde{\phi}\equiv\frac{\phi}{M_p}~,~~~~~\tilde{V}\equiv \frac{V}{\rho_{he0}}~,
 ~~~~~C\equiv\frac{Q^2}{\rho_{he0}M_p^2}~.
 \ee
These two equations relate the hessence potential $V(\phi)$ to the
EOS of the hessence $w_{he}(z)$. Given an effective $w_{he}(z)$,
the construction Eqs.~(\ref{63}) and (\ref{64}) allow us to
construct the hessence potential $V(\phi)$.

 \begin{figure}
 \centerline{\includegraphics[width=10cm]{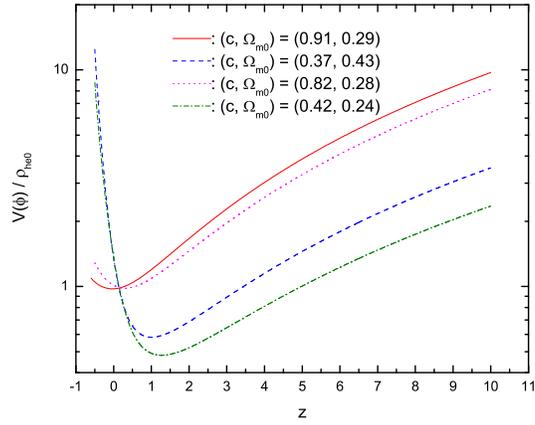}}
 \caption{\small Evolution of potentials of holographic hessence models. }
 \end{figure}

 \begin{figure}
 \centerline{\includegraphics[width=10cm]{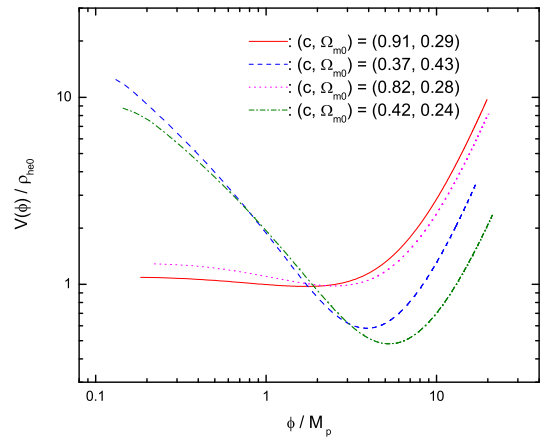}}
 \caption{\small Constructed potentials of holographic hessence models. }
 \end{figure}

Using the solved EOSs of the holographic hessence models in Figure
2, we numerically solve the equations (\ref{63}) and (\ref{64}),
which are shown in Figures 4 and 5, where we have chosen the
initial condition with $C=10.0$ and $\phi_0=0.1$. From Figure 4,
we find that, for the cases with $c=0.37$ and $c=0.42$, the
potentials are decreasing with the expansion of the universe in
the earlier stage, which are same with the quintessence
models\cite{guo}. But after the time, where $z\sim1$, the
potentials begin to increase, and at present time, the potential
functions are increasing functions, which are similar to a phantom
model. For the cases of $c=0.91$ and $c=0.82$, although the
potential functions are monotonically decreasing until the present
time, they all will begin to increase with time in the near
future. Once again, from Figure 5, we find that these four
potentials are all not the monotonic functions, since we have used
$c<1$, which is consistent to the previous analysis. The only
difference is that the lowest positions of these potentials are
different, which is determined by the initial condition and the
parameter $c$. These are different from the simple quintessence,
k-essence and tachyon models\cite{guo}.

\section{summary}

In this letter, we have investigated the hessence models, which
satisfy the holographic principle. The potential of the hessence
is determined by the holographic principle. We have discussed the
evolution of the holographic hessence in the $w-w'$ plane, and
found that the potential function only depends on the parameter
$c$. If $c\geq1$ is chosen, $w_{he}\geq-1$ is kept for all time,
and hessence field $\phi$ rolls down the potential, which is
similar to the quintessence models. However if $c<1$ is chosen,
which mildly favor the observations, the EOS of the models evolve
from the region of $w_{he}>-1$ to that of $w_{he}<-1$, and state
of $w_{he}=-1$ must be crossed. The potential of model is not a
monotonic function. In the early time, the hessence is
quintessence-like, the EOS is $w_{he}>-1$ and $\dot{V}<0$. Then it
enters into the region with $w_{he}>-1$ and $\dot{V}>0$, where the
potential begins to be climbed up. At last, the model must enter
and stay in the phantom-like region with $w_{he}<-1$ and
$\dot{V}>0$. Considered the current constraint of the parameter
$c$, we have reconstruct the potential of the holographic hessence
models, which are all the nonmonotonic functions.

~

\textbf{ACKNOWLEDGMENT}: The author thanks X.Zhang for helpful
discussion.


\newpage


\baselineskip=12truept

\begin{thebibliography}{99}

\bibitem{observation}
A.G.Riess {\it et al.}, Astron.J. {\bf 116}, 1009
(1998);

S.Perlmutter {\it et al.}, Astrophys.J. {\bf 517}, 565
(1999);

C.L.Bennett \emph{et al.}, Astrophys.J.Suppl. {\bf 148}, 1
(2003);

D.N.Spergel \emph{et al.}, arXiv:astro-ph/0603449;

M.Tegmark \emph{et al.}, Astrophys.J. {\bf 606}, 702 (2004);


\bibitem{quint}
C.Wetterich, Nucl.Phys.B {\bf 302}, 668 (1988);

Astron.Astrophys. {\bf 301}, 321 (1995);

B.Ratra and P.J.E.Peebles, Phys.Rev.D {\bf 37}, 3406 (1988);

R.R.Caldwell, R.Dave and P.J.Steinhardt, Phys.Rev.Lett. {\bf 80},
1582 (1998);

W.Zhao, arXiv:astro-ph/0604459;






\bibitem{phantom}
R.R.Caldwell, Phys.Lett.B {\bf 545}, 23 (2002);

S.M.Carroll, M.Hoffman and M.Trodden, Phys.Rev.D {\bf 68}, 023509
(2003);

R.R.Caldwell, M.Kamionkowski and N.N.Weinberg, Phys.Rev.Lett. {\bf
91},  071301 (2003);

M.P.Dabrowski, T.Stachowiak and M.Szydlowski, Phys.Rev.D {\bf 68},
103519 (2003);

V.K.Onemli and R.P.Woodard, Phys.Rev.D {\bf 70}, 107301 (2004);



\bibitem{k}
C.Armendariz-Picon, T.Damour and V.Mukhanov,  Phys.Lett.B {\bf
458}, 209 (1999) ;

T.Chiba, T.Okabe and M.Yamaguchi, Phys.Rev.D {\bf 62}, 023511
(2000);

C.Armendariz-Picon, V.Mukhanov and P.J.Steinhardt, Phys.Rev.D {\bf
63}, 103510 (2001);

T.Chiba, Phys.Rev.D {\bf 66}, 063514 (2002);


\bibitem{quintom}
B.Feng, X.L.Wang and X.M.Zhang, Phys.Lett.B {\bf 607} (2005) 35;

Z.K.Guo, Y.S.Piao, X.M.Zhang and Y.Z~Zhang, Phys.Lett.B {\bf 608},
177 (2005);

X.F.Zhang, H.Li, Y.S.Piao and X.M.Zhang, Mod.Phys.Lett.A {\bf 21},
231 (2006);

M.Z.Li, B.Feng and X.M.Zhang, JCAP {\bf 0512}, 002 (2005);



\bibitem{vector}
W.Zhao and Y.Zhang, Class.Quant.Grav. {\bf 23}, 3405 (2006);

W.Zhao and Y.Zhang, Phys.Lett.B {\bf 640}, 69 (2006);

W.Zhao and D.H.Xu, Int.J.Mod.Phys.D accepted
(arXiv:gr-qc/0701136);

Y.Zhang, T.Y.Xia and W.Zhao, arXiv:gr-qc/0609115;



\bibitem{gt}
G.Dvali, G.Gabadadze and M.Porrati, Phys.Lett.B {\bf 485}, 208
(2000);

G.Dvalli and G.Gabadadze, Phys,Rev.D {\bf 63}, 065007 (2001);

H.S.Zhang and Z.H.Zhu, Phys.Rev.D {\bf 75}, 023510 (2007);

\bibitem{Hsu:2004ri}
S.D.H.Hsu, Phys.Lett.B {\bf 594}, 13 (2004);



\bibitem{Li:2004rb}
M.Li, Phys.Lett.B {\bf 603}, 1 (2004);




\bibitem{cohen}
A.Cohen, D.Kaplan and A.Nelson, Phys.Rev.Lett.{\bf 82}, 4971
(1999);



\bibitem{c>1}
J.Y.Shen, B.Wang, E.Abdalla and R.K.Su, Phys.lett.B {\bf609}, 200
(2005);

K.Enquivst and M.S.Sloth, Phys.Rev.Lett. {\bf 93}, 221302 (2004);

K.Enqvist, S.Hannestad and M.S.Sloth, JCAP {\bf 0502}, 004 (2005);

X.Zhang, Int.J.Mod.Phys.D {\bf 14}, 1597 (2005);



\bibitem{-1}
A.Vikman, Phys.Rev.D {\bf 71}, 023515 (2005);

\bibitem{hessence}
H.Wei, R.G.Cai and D.F.Zeng, Class.Quant.Grav. {\bf 22}, 3189
(2005);

H.Wei and R.G.Cai, Phys.Rev.D {\bf 72}, 123507
(2005);

M.Alimohammadi and H.M.Sadjadi, Phys.Rev.D {\bf 73}, 083527
(2006);

H.Wei, N.Tang and S.N.Zhang, Phys.Rev.D {\bf 75} 043009 (2007);

H.Wei and S.N.Zhang, arXiv:0704.3330v2;

\bibitem{zhao}
W.Zhao and Y.Zhang, Phys.Rev.D {\bf 73}, 123509 (2006);

\bibitem{brane}
I.Ya.Aref'eva, A.S.Koshelev and S.Yu.Vernov, Phys.Rev.D {\bf 72}
(2005) 064017;




\bibitem{snap}
SNAP Collaboration, arXiv:astro-ph/0507458;

SNAP Collaboration, arXiv:astro-ph/0507459;


\bibitem{obs1}
X.Zhang and F.Q.Wu, Phys.Rev.D {\bf 72}, 043524 (2005);

Z.Chang, F.Q.Wu and X.Zhang, Phys.Lett.B {\bf 633}, 14 (2006);

Q.G.Huang and Y.G.Gong, JCAP {\bf 0408}, 006 (2004);

K.Enqvist, S.Hannestad and M.S.Sloth, JCAP {\bf 0502} 004 (2005);

J.Shen, B.Wang, E.Abdalla and R.K.Su, Phys.Lett.B {\bf 609}, 200
(2005);

H.C.Kao, W.L.Lee and F.L.Lin, Phys.Rev.D {\bf 71}, 123518 (2005);


\bibitem{wwp}
R.R.Caldwell and E.V.Linder, Phys.Rev.Lett. {\bf 95}, 141301
(2005);

R.J.Scherrer, Phys.Rev.D {\bf 73}, 043502 (2006);

T.Chiba, Phys.Rev.D {\bf 73}, 063501 (2006);





\bibitem{zx}
X.Zhang, Phys.Rev.D {\bf 74}, 103505 (2006);

X.Zhang and F.Q.Wu, arXiv:astro-ph/0701405;

\bibitem{snia}
A.G.Riess \emph{et al.}, arXiv:astro-ph/0611572;


\bibitem{seljak}
U.Seljak, A.Slosar and P.McDonald, JCAP {\bf 0610}, 014 (2006);

\bibitem{cmb}
Y.Wang and P.Mukherjee, Astrophys.J. {\bf 650}, 1 (2006);

\bibitem{bao}
D.J.Eisenstein \emph{et al.}, Astrophys.J. {\bf 633}, 560 (2005);

\bibitem{h}
W.L.Freeman \emph{et al.}, Astrophys.J. {\bf 553}, 47 (2001);


\bibitem{0506310}

X.Zhang and F.Q.Wu, Phys.Rev.D {\bf 72} 043524 (2005);




\bibitem{guo}
Z.K.Guo, N.Ohta and Y.Z.Zhang, Phys.Rev.D {\bf 72}, 023504
(2005);

H.Li, Z.K.Guo and Y.Z.Zhang, Mod.Phys.Lett.A {\bf 21}, 1683
(2006);



\end{thebibliography}
\end{document}